\documentclass[iop]{emulateapj}
\usepackage[]{times,url,graphicx,amssymb,epsfig,longtable,subfigure,ulem}

\usepackage{xspace}

\newcommand{\water}{H$_{2}$O\xspace}

\newcommand{\twc}{$^{12}$CO\xspace}
\newcommand{\thc}{$^{13}$CO\xspace}
\def\Msun{M$_{\sun}$\xspace}
\def\Lsun{L$_{\sun}$\xspace}

\def\kms{km s$^{-1}$\xspace}
\def\fLowJ{0.14\xspace}
\def\vShock{30\xspace ${\rm km/s}$\xspace}
\def\nShock{$2\times 10^{4}$~$\rm{cm^{-3}}$\xspace}

\def\G0{}
\accepted{11/15/2013 for publication in ApJL}
\begin{document}
\slugcomment{}
\shorttitle{A case for shocked molecules in NGC~1266}
\shortauthors{Pellegrini et al.}

\title{ Shock Excited Molecules in NGC~1266: ULIRG conditions at the center of a
  bulge-dominated galaxy}
\author{E.~W.~Pellegrini$^1$,J.~D.~Smith(PI)$^1$, M.~G.~Wolfire$^2$,
  B.~T.~Draine$^3$, A.~F.~Crocker$^1$, K.~V.~Croxall$^4$, P.~van der
  Werf$^5$, D.~A.~Dale$^6$, D.~Rigopoulou$^{7,8}$,
  C.~D.~Wilson$^{9}$, E.~Schinnerer$^{10}$, B.~A.~Groves$^{10}$,
  K.~Kreckel$^{10}$, K.~M.~Sandstrom$^{10}$, L.~Armus$^{11}$,
  D.~Calzetti$^{12}$, E.~J.~Murphy$^{11}$, F.~Walter$^{10}$,
  J.~Koda$^{13}$, E.~Bayet$^{14}$, P.~Beirao$^{15}$,
  A.~D.~Bolatto$^2$, M.~Bradford$^{11}$, E.~Brinks$^{16}$,
  L.~Hunt$^{17}$, R.~Kennicutt$^{18}$, J.~H.~Knapen$^{19,20}$,
  A.~K.~Leroy$^{21}$, E.~Rosolowsky$^{22}$, L.~Vigroux$^{23}$,
  R.~H.~B.~Hopwood$^{24}$}

\affil{
$^1$Department of Physics and Astronomy, University of Toledo, Toledo, OH 43606, USA,
$^2$ Department of Astronomy, University of Maryland, College Park, MD 20742, USA,
$^3$Princeton University Observatory, Peyton Hall, Princeton, NJ 08544-1001, USA,
$^4$Department of Astronomy, The Ohio State University, 140 West 18th Avenue, Columbus, OH 43210, USA,
$^5$Leiden Observatory, Leiden University, P.O. Box 9513, 2300 RA Leiden, The Netherlands,
$^6$Department of Physics and Astronomy, University of Wyoming, Laramie, WY 82071, USA 
$^7$Department of Astrophysics, University of Oxford, Keble Road, Oxford, OX1 3RH, UK,
$^{8}$RAL Space, Rutherford Appleton Laboratory, Chilton, Didcot OX110QX, UK,
$^{9}$Department of Physics \& Astronomy, McMaster University, Hamilton, Ontario L8S 4M1, Canada,
$^{10}$Max Planck Institut fur Astronomie, Konigstuhl 17, 69117 Heidelberg, Germany,
$^{11}$ Spitzer Science Center, California Institute of Technology, MC 314-6, Pasadena, CA 91125, USA,
$^{12}$Department of Astronomy, University of Massachusetts, Amherst, MA 01003, USA,
$^{13}$Department of Physics and Astronomy, SUNY Stony Brook, Stony Brook, NY 11794-3800, USA,
$^{14}$Department of Physics and Astronomy, University College London, Gower Street, London WC1E 6BT, UK,
$^{15}$Observatoire de Paris, 61 avenue de l'Observatoire, Paris, 75014, France,
$^{16}$Centre for Astrophysics Research, University of Hertfordshire, Hatfield AL 10 9AB, UK,
$^{17}$INAF-Osservatorio Astrosico di Arcetri, Largo E. Fermi 5, I-50125 Firenze, Italy,
$^{18}$Institute of Astronomy, University of Cambridge, Madingley Road, Cambridge CB3 0HA, UK,
$^{19}$Instituto de Astrof´ısica de Canarias, 38200 La Laguna, Spain,
$^{20}$Departamento de Astrof\'\i sica, Universidad de La Laguna, E-38206 La Laguna, Spain
$^{21}$National Radio Astronomy Observatory, 520 Edgemont Road, Charlottesville, VA 22903, USA,
$^{22}$Department of Physics, University of Alberta, 2-115 Centennial Centre for Interdisc Science, Edmonton, AB, Canada,
$^{23}$Institut d’Astrophysique de Paris, UMR 7095 CNRS, Universit´e Pierre et Marie Curie, 75014 Paris, France
$^{24}$Department of Physics and Astronomy, Open University, Walton Hall, Milton Keynes MK7 6AA, UK
}
\email{eric.pellegrini@utoledo.edu}


\begin{abstract}
We investigate the far infrared spectrum of NGC~1266, a S0 galaxy that
contains a massive reservoir of highly excited molecular gas. Using
the SPIRE-FTS, we detect the $^{12}$CO ladder up to J=(13-12), [\ion{C}{1}]
and [\ion{N}{2}] lines, and also strong water lines more characteristic of
UltraLuminous IR Galaxies (ULIRGs). The $^{12}$CO line emission is
modeled with a combination of a low-velocity C-shock and a PDR. Shocks
are required to produce the H$_2$O and most of the high-J CO
emission. Despite having an infrared luminosity thirty times less than
a typical ULIRG, the spectral characteristics and physical conditions
of the ISM of NGC~1266 closely resemble those of ULIRGs, which often
harbor strong shocks and large-scale outflows.
\end{abstract}


\section{Introduction}
\label{sec:Introduction}
The IR emission of galaxies beyond the peak of their thermal dust
emission, from 200-600~$\micron$, traces their ionized, neutral and
molecular interstellar medium (ISM). In particular the \twc
transitions (hereafter CO) from J=4-3 to J=13-12 provide an important
diagnostic of molecular gas excitation. These CO spectral line
energy distributions (SLEDs) allow one to distinguish those regions
within galaxies dominated by star formation (SF)(photo dissociation
regions, PDRs) or by AGN produced X-ray dominated regions (XDRs).

At a distance of 29.9~Mpc (z=0.0073), NGC~1266 contains $1.1 \times
10^{9}$~\Msun of molecular gas within a 100 pc radius, with a denser,
60~pc core containing $4\times 10^{8}$~\Msun of gas
\citep{Alatalo2011}. In contrast, nearby giant molecular clouds of a
similar size contain only $10^{6}$ \Msun of gas \citep{Fukui2010}. The
core mass and projected size corresponds to a surface density of $2.7
\times 10^{4}$ \Msun pc$^{-2}$, similar to what is observed in ULIRGs
\citep{Bothwell2010}. Its IRAS 60/100$\micron$ ratio is twice as large
as the median value of more ordinary galaxies within the SINGS sample
\citep{Dale2009}, yet within 5\% of the mean ratio of 41 ULIRGs
\citep{Farrah2003}. The observed 9.7$\micron$ silicate optical
  depth $\tau({\rm Si}) \approx 2$ ({\it Spitzer} IRS spectra;
  \citealt{Smith2007}).  The ISM column density and dust colors are
  quite comparable with samples of ULIRGS, despite NGC~1266's total
  infrared luminosity (TIR) of only $ 2.75~\times~10^{10}$~\Lsun.

Broad velocity wings of the low-J CO lines reveal a massive and
energetic molecular outflow ($M_{\rm out} = 2.4 \times 10^7$~\Msun,
$\dot{M}_{\rm out} = 13$~\Msun/yr; \citealt{Alatalo2011}). The mass
outflow rate exceeds the SF rate by a factor of 5, leading to a
highly obscured, weak AGN as a likely driving mechanism
\citep{Alatalo2011}. Evidence of mechanical feedback is also detected
in optical lines from ionized gas, with shock velocities $\approx 500$
\kms \citep{Davis2012}.

\section{Observations}
Our observations include Herschel Fourier Transform Spectrometer
(SPIRE-FTS) spectroscopy obtained as part of the ``Beyond The Peak''
(BTP; OT1\_jsmith1; PI J.D. Smith) survey of 21 nuclear regions and 2
extra-nuclear regions selected from the SINGS \citep{Kennicutt2003}
and KINGFISH \citep{KINGFISH} surveys. BTP provides deep, intermediate
spaced mapping of a large collection of nearby galaxies at $^{12}$CO
transitions above J=(3-2).

\subsection{PACS spectroscopy} 
NGC~1266 has been imaged in [\ion{C}{2}], [\ion{O}{1}], [\ion{O}{3}],
and [\ion{N}{2}] emission lines as part the Herschel Key Program
KINGFISH \citep{KINGFISH}.  The data were using the KINGFISH pipeline
for PACS spectroscopy in HIPE v9.0.3063 using the calibration files in
PACS calVersion 41, following the methods described in
\citet{Croxall2012}.

\subsection{SPIRE-FTS spectroscopy}
All SPIRE-FTS observations use a 4-point dither which sub-critically
samples extended emission with an internal jiggle mirror. Our data
are calibrated with HIPE v10 using an extended-flux
calibration. The default pipeline includes a central, unjiggled
telescope response function (RSRF) created from thousands of
repetitions of dark observations. We calculate new jiggled RSRFs from
$\simeq 100$ repetitions of jiggled darks, now available in
HIPE~v11. For dark subtraction, we use the un-jiggled deep darks
obtained during the same FTS observing run and fit a high-order
polynomial to each bolometer.

While NGC~1266 is a compact source, the SLW bolometer beam size and
dither pattern produce repeat, but off-center SLW observations of our
target.  At this time, dithered calibrations have only been obtained
for extended sources. To calibrate these dithered observations we
begin with the Jiggle-0 (undithered) spectrum, processed with the
point-source pipeline. Next all data were re-processed as with mapped
calibrations, and an average spectrum was created by weighting
individual SLW bolometers with the projected overlap of the beam
profile and a 20'' aperture centered on the peak of the SPIRE
250$\micron$ image \citep{KINGFISH}. We ratio the mapped and point
source spectra to create a transfer function to calibrate the dithered
observations while simultaneously accounting for the beam filling
factor of a point source. Off target bolometers also sample ample sky
around NGC~1266 and we find no continuum or line emission, and thus
perform no background subtraction.

\begin{figure*}\begin{center}
  \includegraphics[scale=0.8]{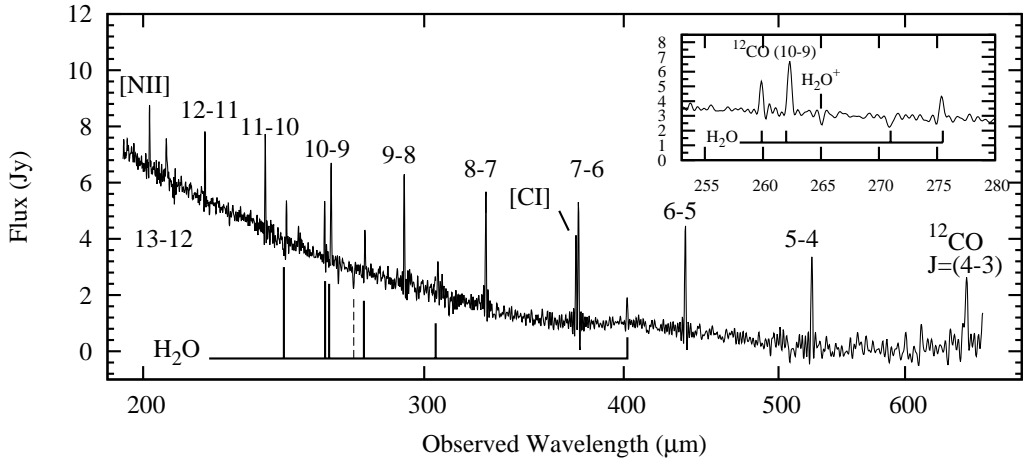}
\caption{\footnotesize The SPIRE-FTS spectrum from 200 to
  600~$\micron$. Emission lines include $^{12}$CO from J=(4-3) to
  J=(13-12), [\ion{N}{2}] 205, [\ion{C}{1}] 370 and
  609~$\micron$. Strong ortho and para \water emission characteristic
  of ULIRGS is seen in emission (solid line) as well as absorption
  (dashed line).}
\label{fig:SPIRESpec}
\end{center}
\end{figure*}

The resulting spectrum (Figure \ref{fig:SPIRESpec}) shows prominent
lines of [\ion{N}{2}], [\ion{C}{1}], 10 transitions of \twc, and 7
transitions of \water. We measure emission line fluxes from the
unapodized spectrum using sinc profiles with a variable spectral
width. Due to outflows we allow line central velocities to vary up to
100 km/s from the systematic velocity. The line fluxes and errors
associated with the continuum uncertainty are listed in Table
\ref{tab:Fluxes}, which also includes ground based measurements of
\twc J=(1-0), (2-1) and (3-2) from \citet{Alatalo2011}. In addition to the
H$_2$ lines from Spitzer IRS, we also add NIR 1-0 S(1) H$_2$ fluxes
from \citet{Riffel2013}.

Table \ref{tab:Fluxes} line fluxes are extinction corrected using
the PAHFIT fully-mixed geometry decomposition of NGC~1266's low
resolution IRS spectroscopy, which yielded a silicate opacity
$\tau_{9.7\mu m} = 2.05$. The correction if we adopt a screen geometry
is negligibly different.

\section{Powering Molecular Emission}
We consider energy sources that are capable of powering the emission
from the central 100~pc of NGC~1266: an AGN and SF. X-ray emission
from an obscured AGN could produce an XDR, while star-formation would
produce a PDR. Both phenomena are also capable of driving shocks,
which would mechanically heat the gas. However, we emphasize that the
dominance of one heating mechanism does not preclude the presence of
the other.

\subsection{CO and H2}
\subsubsection{Shocks}
Motivated by shocks detected in optical lines, and molecular outflows, we
first consider shocked gas as the source of CO excitation. To fit the CO
SLED, we form combinations of the J- and/or C-shock models from
\citep{Flower2010}. We then fit these to observations after
normalization by total intensity from J=(1-0) to (13-12) according to
\begin{equation}
I_{norm}(J_{i}) = \sum_{k} f_{k}({\rm ^{12}CO}) \times
\frac{I_{k}(J_i)}{\sum_{i=1}^{13} I_{k}(J_{i})}
\end{equation}
where the $k^{th}$ model contributes a fraction $f_{k}$ to the total
predicted CO intensity, with a minimum $f_{k}$=0.01 and $ \sum_{k}
f_{k} = 1$. 

With a large reservoir of dense gas we expect some level of
star-formation in NGC~1266. The observed TIR emission is consistent
with SF$\approx$~2.1\Msun/yr \citep{Alatalo2011}. So we consider
combinations of shock and plane-parallel PDR models.  To estimate the
emission from PDRs, we adopt the models of \citet{Wolfire2010}, with
updates in \citet{Hollenbach2012}. Note that we adopted the shock and
PDR models with solar abundance as calculated. We do not include the
60/100$\micron$ ratio from the dust SED as a constraint due to
uncertainties in the optical depth of the dust emission. However, we
do constrain $\beta_{CO} = L_{CO}/L_{\rm TIR}$ not to exceed the
observed ratio, based upon the assumption that shocks do not contribute
appreciably to $L_{\rm TIR}$ and that a single face-on PDR contributes
a continuum with an intensity related to the far-UV radiation field
$G_0$ as
\begin{equation} 
I_{\rm TIR}(G_{\rm 0}) \simeq \frac{3\times1.6 \times 10^{-3}}{4\pi} \times G_{\rm 0} ({\rm erg~s^{-1}~cm^{-2}~sr^{-1}})
\end{equation}.

A PDR fits the lowest-J CO lines with $n({\rm
  H})=10^{3.5}$~${\rm cm^{-3}}$, ${\rm log}~ G_{0}=2.25$, while
C-shocks fit the higher-J lines with $v=30$ ${\rm km ~s^{-1}}$,
$n({\rm H}) = 2\times 10^{4}$~${\rm cm^{-3}}$ (reduced $\chi^2 =
2.6$). The best fitting combination of the existing models are shown
in Figure \ref{fig:COSLED}a with a solid line (black), as well as the
individual PDR~(blue) and shock~(red) components. Model parameter
probability distribution functions (PDFs) are shown on the right.
  The NGC~1266 \water SLED is also shown top-right. Although it was
  not used as an input, the best-fit model reproduces it well,
  including the faint high temperature transitions. The shock
under-predicts the [\ion{O}{1}]~63$\micron$, but the [\ion{O}{1}] intensity varies
strongly with shock parameters. A small contribution from a 10~km/s
shock we would explain the observed [O~I] intensity.

\subsubsection{PDR+PDR}
 A dual PDR model ($n_1{\rm(H)}=10^{3.75} {\rm cm^{-3}}$, $G_{\rm
   0,1}~=~10^{6}$; $n_2{\rm(H)}=10^{5.5}$, $G_{\rm 0,2} = 10^{3.5}$)
 could equally well explain the \twc and H$_2$ excitation and relative
 intensity. We reject the dual-PDR scenario because it requires ${\rm
   G_0}\approx 10^4 - 10^6$, resulting in $\beta_{CO}~1000\times$
 lower than that observed.  \citet{Meijerink2013} similarly rejected a
 dual PDR in NGC~6240 based on this argument. In NGC~1266, a dual PDR
 under-predicts $\beta_{CO}$, and hence over-predicts $L_{\rm
   TIR}$, by 3 orders of magnitude. With $\beta$ constrained to
 $\ge~4.6~\times~10^{-4}$, the resulting reduced $\chi^2$ is $\ge
 170$. Thus a two component PDRs is rejected as the sole source of the
 molecular emission.

\subsubsection{XDR}
The ROSAT X-ray flux of $2.49\times10^{-13}\pm0.3~{\rm erg~s^{-1}
  cm^{-2}}$ from 0.2-2.0~keV \citep{WGACAT} could also indicate an XDR
contribution to \twc and the \water lines detected in our spectrum. We
will fully explore the possible contribution of XDRs in our full
survey, which includes 10 AGN, in a future paper. However, an XDR is
unlikely to dominate the \twc emission in NGC~1266.  First, XDRs
produce high ionization fractions in molecular gas leading to
${\rm H_{2}O^+}$ and OH$^+$ emission with comparable intensities to \water
lines (\citealt{vanderWerf2010,Meijerink2013}), yet no emission from
these molecules is detected. We do detect absorption from
$\rm{H_2O^+}$, but this has been seen in systems where XDR are ruled
out (e.g. Arp220, \citealt{Rangwala2011}).

\begin{figure*}
\begin{center}
\includegraphics[scale=0.75]{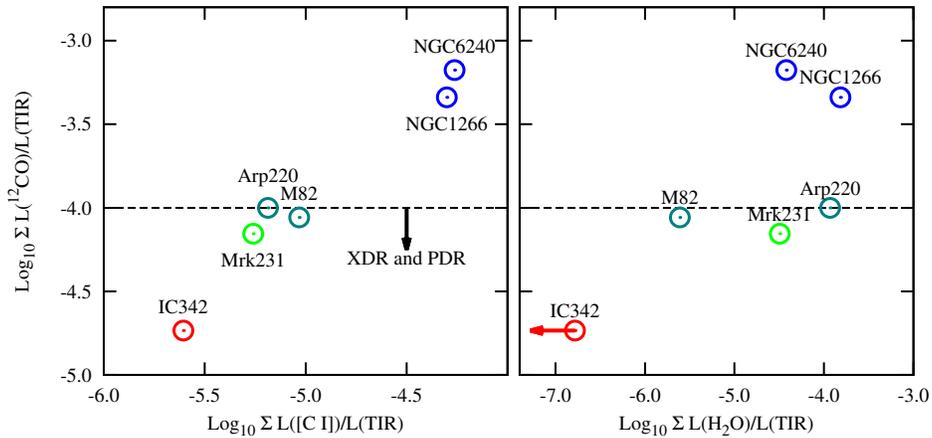}
\caption{Left: Total CO, from J=(1-0) to (13-12), relative to TIR, and
  [CI]$_{370,609 \micron }$ relative to TIR for galaxies with detected
  \water emission and a detailed analysis of gas excitation
  mechanisms. Galaxies are color coded as PDRs (red), XDRs (green),
  mechanical (teal) and shocks (blue). Right: Same as left, with total
  L(\water)/TIR on the x-axis. The horizontal dashed line marks the
  upper limit of L(CO)/L(TIR) for XDRs and PDRs
  (\citealt{Meijerink2005,Meijerink2007}).}
\label{fig:TIRCO}
\end{center}
\end{figure*}

Second, $\beta_{CO}$ also eliminates XDRs as the dominant heating
source in NGC~1266.  The efficiency at which X-rays are converted to
line emission and IR continuum is very different than in shocks which
heat gas but are relatively inefficient at heating dust. This is
illustrated in Figure~\ref{fig:TIRCO}, a plot of $\beta_{CO}$ against
$\beta_{[CI]}$ and $\beta_{H_2O}$. As discussed in
\citet{Meijerink2013}, XDR and PDR models have an upper limit on
the line-to-continuum ratio of $\beta_{\rm CO} \le 10^{-4}$,
represented by the dashed line in Figure~\ref{fig:TIRCO}. In NGC~1266
that ratio for the 13 detected CO lines is nearly $10^{-3}$, just
below that of NGC~6240 (see
Figure~\ref{fig:TIRCO}). \citet{Alatalo2011} found that the observed
$\beta_{\rm H_{2}}$ ratio was 3 times larger than could explained by
an XDR.  The ratio of $\beta_{[CI]}$ is also enhanced by an order of
magnitude in NGC~1266, as in NGC~6240.  The linearity of [\ion{C}{1}] with
\twc, seen in Figure~\ref{fig:TIRCO}, suggests it is a powerful
diagnostic especially useful where the entire \twc is not observable,
as is the case with high redshift galaxies observed with ALMA.

\begin{figure*}
  \begin{center}
    \begin{tabular}{c}
      \includegraphics[scale=0.6]{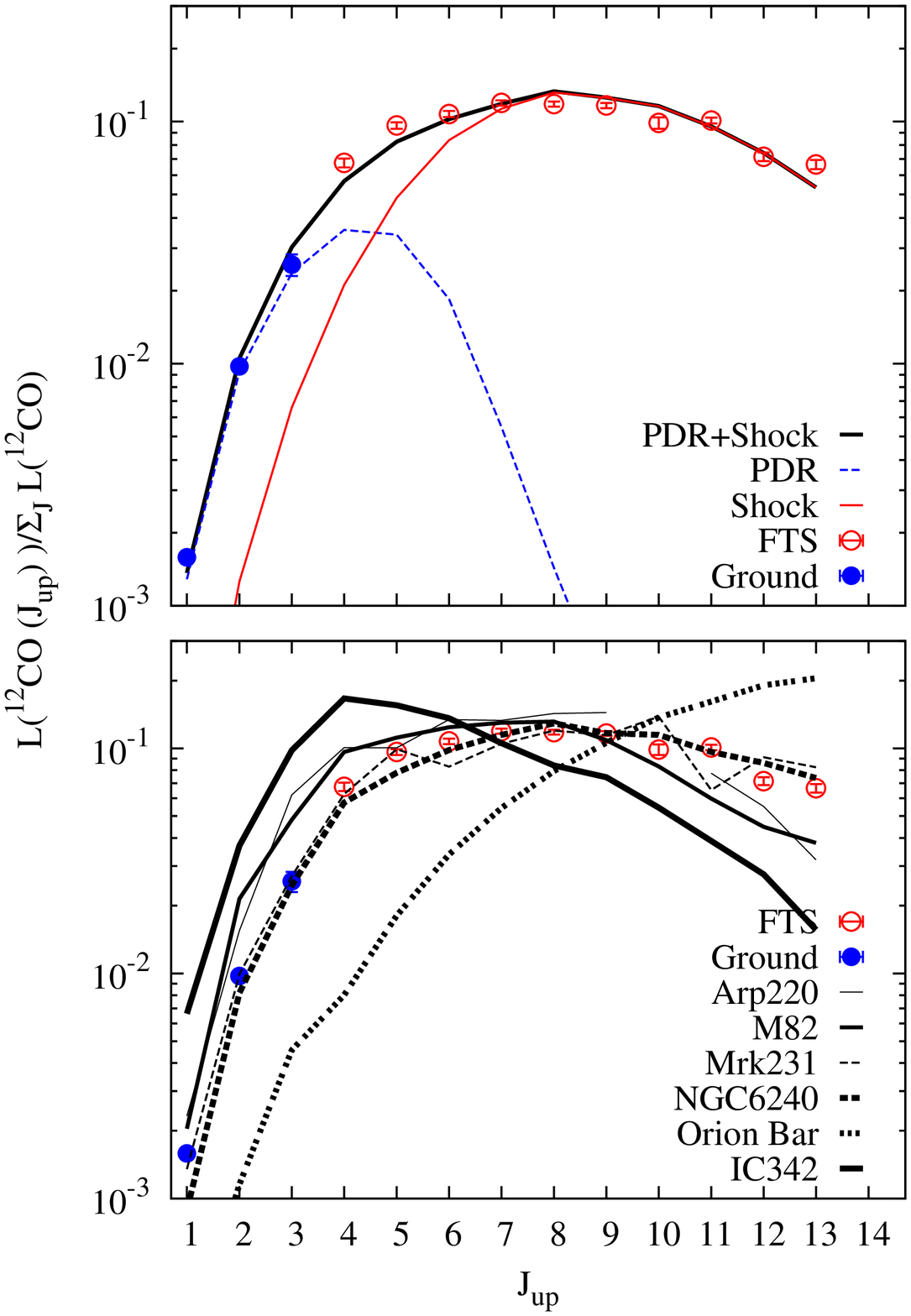}
      \includegraphics[scale=0.59]{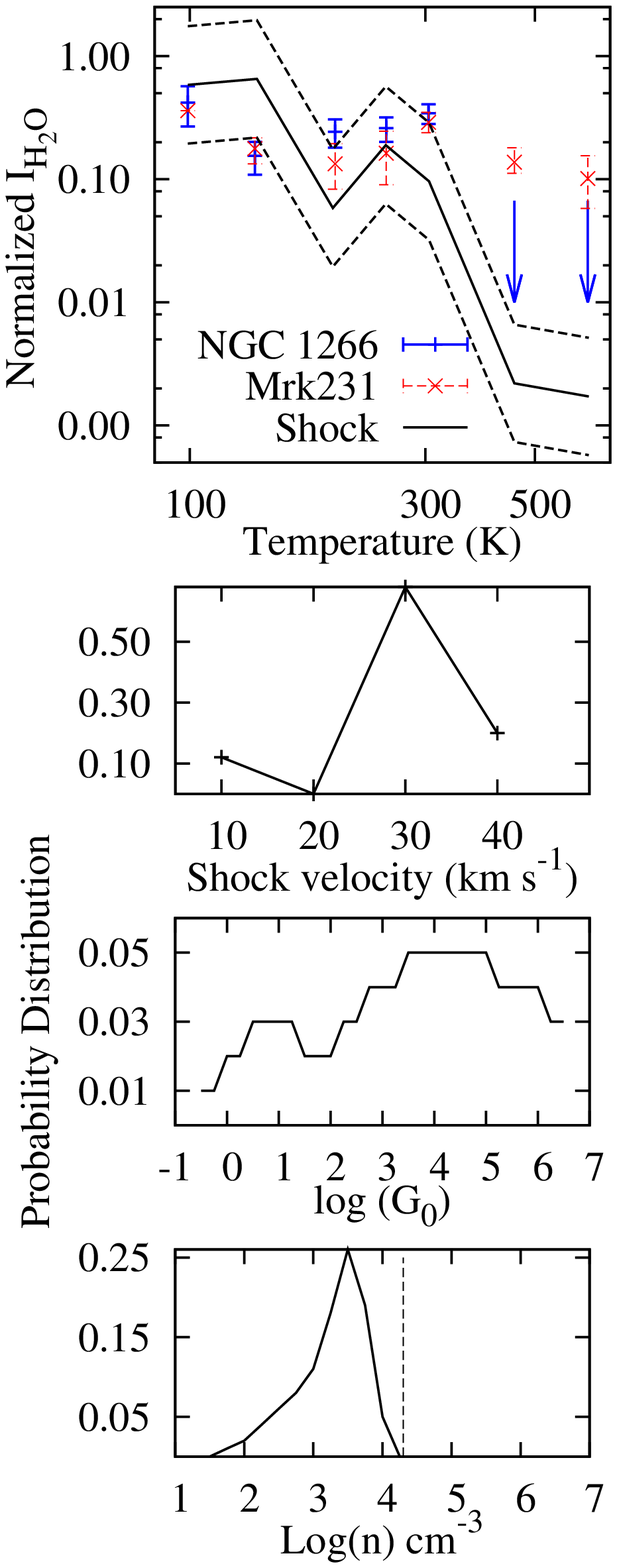}
    \end{tabular}
  \end{center}
  \caption{(Top-Left) NGC~1266 $^{12}$CO SLED (up to J=(13-12))
  (circles) normalized by total observed $^{12}$CO luminosity. Shown
  is the best fitting PDR (dashed blue) and Shock model (solid
  red). (Bottom-Left) Comparison of $^{12}$CO SLEDs of various
  galaxies, and the Orion Bar.  (Right) The \water SLED includes
  NGC~1266, Mrk231 and the best fit shock model with a factor of 3
  uncertainty in individual lines. Normalized probability
  distributions functions of the PDR+Shock parameters, and the \water
  SLED.  A vertical dashed-line represents the only pre-shock density
  that fits observations. }
  \label{fig:COSLED}
\end{figure*}

\begin{deluxetable}{lcccc}
\tabletypesize{\footnotesize}
\tablecaption{NGC~1266 IRS, PACS and SPIRE-FTS Extinction Corrected Line Fluxes}
\tablehead{\colhead{Species}&\colhead{Trans}&\colhead{Wave($\micron$)}&\colhead{Flux($\rm{10^{-17} W m^{-2}}$)}&\colhead{Inst.}}
\startdata
\twc&1-0&2600&0.063$\pm$0.004&$^1$\\
\twc&2-1&1300&0.39$\pm$0.015&$^1$\\
\twc&3-2&867.0&1.02$\pm$0.11&$^1$\\
\twc&4-3&650.7&2.69$\pm$1.14&$^2$\\
\twc&5-4&520.6&3.84$\pm$0.69&$^2$\\
\twc&6-5&433.9&4.28$\pm$0.35&$^2$\\
\twc&7-6&371.9&4.76$\pm$0.28&$^2$\\
\twc&8-7&325.5&4.71$\pm$0.36&$^2$\\
\twc&9-8&289.3&4.64$\pm$0.65&$^2$\\
\twc&10-9&260.4&3.94$\pm$0.60&$^2$\\
\twc&11-10&236.8&4.03$\pm$0.61&$^2$\\
\twc&12-11&217.1&2.85$\pm$0.30&$^2$\\
\twc&13-12&200.4&2.65$\pm$0.58&$^2$\\
\water&$2_{11}-2_{02}$&398.9&1.09$\pm$0.32&$^2$\\
\water&$2_{02}-1_{11}$&303.7&2.91$\pm$1.04&$^2$\\
\water&$3_{12}-3_{03}$&273.4&1.80$\pm$0.40&$^2$\\
\water&$3_{12}-2_{21}$&260.2&1.83$\pm$0.61&$^2$\\
\water&$3_{21}-3_{12}$&258.0&2.38$\pm$0.45&$^2$\\
\water&$2_{20}-2_{11}$&244.1&1.69$\pm$0.45&$^2$\\
\hline
\water&$4_{22}-4_{13}$&248.2&$<$0.47&$^2$\\
\water&$5_{23}-5_{14}$&212.5&$<$0.47&$^2$\\
\hline
{[}\ion{C}{1}]&\nodata&370.7&3.19$\pm$0.26&$^2$\\
{[}\ion{C}{1}]&\nodata&609.6&1.13$\pm$0.86&$^2$\\
{[}\ion{C}{2}]&\nodata&157.7&17.30$\pm$0.46&$^3$\\
{[}\ion{O}{1}]&\nodata&63.18&11.40$\pm$1.32&$^3$\\
{[}\ion{N}{2}]&\nodata&205&1.70$\pm$0.88&$^2$\\
\hline
${\rm H_2}$&S(4)&8.026&21.90$\pm$1.01&$^4$\\
${\rm H_2}$&S(3)&9.662&13.00$\pm$0.39&$^4$\\
${\rm H_2}$&S(2)&12.282&10.90$\pm$0.26&$^4$\\
${\rm H_2}$&S(1)&17.035&8.48$\pm$0.10&$^4$\\
${\rm H_2}$&S(0)&28.171&2.94$\pm$0.22&$^4$\\
${\rm H_2}$&1-0S(1)&2.1213&10.90$\pm$0.20&$^5$
\enddata

\tablecomments{Ground \twc from $^1$\citet{Alatalo2011}. Lines
  observed with $^2$SPIRE-FTS, $^3$PACS, $^4$Spitzer-IRS
  instruments. NIR H$_2$ from $^5$\citet{Riffel2013}. IRS fluxes have
  been extracted over a 10'' aperture centered on the peak of the IR
  continuum emission which is unresolved by Spitzer and Herschel.}

\label{tab:Fluxes}
\end{deluxetable}

\subsection{Shocked H$_{2}$O}
The gas phase abundance of \water is driven by two processes:
formation via endothermic reactions and the release from ices on the
surface of dust grains. J- and C-shocks, as well as XDRs, are capable
of producing the warm and dense molecular gas characteristic of \water
emitting regions.

\water emission is not included in the \citet{Wolfire2010} PDR models
because it is very faint. Therefore, we assume the Orion Bar value of
L(\water)/L(CO)= 0.001 \citep{Habart2010} for our modeled PDRs. The
observed parameters of the Orion Bar PDR $G_{0}\approx 4\times10^{4}$
\citep{Tielens1985} and $n({\rm H})\approx 10^5$~${\rm cm^{-3}}$
\citep{Allers2005, Pellegrini2009} are of higher density and
excitation than the PDR parameters necessary to explain the low
excitation CO in NGC~1266. These energetic conditions favor the
formation of water, and thus provide an upper limit to the \water
emission we expect from the PDR.

Table \ref{tab:H2O} and Figure\ref{fig:TIRCO} summarize the ratio of
\water, \twc and [CI] to L(TIR). We examine the ratio of \water to
\twc because both originate in molecular gas, unlike the IR
continuum. The Shock+PDR combination predicts a L(\water)/L(CO)=0.34,
very similar to the NGC~1266 value of 0.33. The observed value is also
similar to that found in the XDR heated Mrk231 \citep{vanderWerf2010},
and the mechanically heated Arp220 \citep{Rangwala2011}. Thus, total
L(\water) relative to L(CO) or L(TIR) is limited as a shock diagnostic
because it can be enhanced by radiative pumping in systems with high
gas column densities, resulting in bright emission from high levels of
\water. However, in a sample of IR pumped galaxies, typical \water
transitions are \water$2_{02}-1_{11}/4_{22}-4_{13} \leq 2.5$ (Yang et
al., in prep). In Figure~\ref{fig:COSLED} \water $4_{22}-4_{13}$ and
$5_{23} - 5_{14}$ are undetected in NGC~1266, with
$2_{02}-1_{11}/4_{22}-4_{13} \geq 4.0$. Thus, the intense emission and
weak excitation of \water emission provides strong evidence for shock
heating of molecular gas in NGC~1266.

\begin{deluxetable*}{lccccc}
\tabletypesize{\footnotesize}
\tablecaption{Line to Continuum Ratios}
\tablehead{\colhead{Object}&\colhead{L(TIR)}&\colhead{L(CO)/L(TIR)}&\colhead{L(\water)/L(TIR)}&\colhead{L([CI])/L(TIR)}&\colhead{Heating}\\
           \colhead{}      &\colhead{$\times 10^{10}$}&\colhead{$\times 10^{-5}$}&\colhead{$\times 10^{-7}$}&\colhead{$\times 10^{-6}$}&\colhead{}}
\startdata
\tablenotemark{a}IC342    &1.3  &1.77&1.59&2.5& PDR\\
\tablenotemark{b}M82      &5.6  &8.75&24.5&9.3& Mech\\
\tablenotemark{c}NGC~6240 &75   &66.7&380&54.7& Shock\\
\tablenotemark{d}{\bf NGC~1266} & {\bf 2.4} &{\bf 45.8}&{\bf 1534}&{\bf 50.0}&{\bf Shock}\\
\tablenotemark{e}Mrk 231  &400  &7.00&321& 5.5& XDR\\
\tablenotemark{f}Arp220   &200  &10.0&1170& 6.5& Mech\\
\enddata

\tablecomments{References: $^{a}$\citet{Rigopoulou2013},
  $^{b}$\citet{Kamenetzky2012}, $^{c}$\citet{Meijerink2013},$^{d}$This
  work, $^{e}$\citet{vanderWerf2010},
  $^{f}$\citet{Rangwala2011}.}

\label{tab:H2O}
\end{deluxetable*}

\section{Discussion}
\subsection{Driving Shocks: Star-formation or an AGN}
The shock model which fits the CO SLED converts 70\% of incident
mechanical power into observable H$_{2}$ emission. Taking the observed
$L(H_{2}) = 1.5 \times 10^{7}$ \Lsun, this requires a incident
mechanical luminosity of $2.1\times 10^7$ \Lsun.

\citet{Alatalo2011} argued for an AGN based on the ratio of mass
outflow to SF rate of $\dot{M}_{out}/\dot{M}_{SF} = 5$. However,
recent models used to explain massive outflows in high redshift
galaxies find SF driven outflow rates between 1-8$\times$SF rate
(Bournaud et al., 2013), suggesting SF could dominate. To determine
the possible contribution from SF we assume the core SF rate is
2.1\Msun/yr \citep{Alatalo2011}. From \citet{Hopkins2011}, the energy
injection rate into a turbulent medium is
\begin{equation}
\dot{E} = (1+\eta_{p} \tau)\times 4\times 10^{-4}\times SF \times {\rm c} \delta v
\end{equation}.

We assume radiation and winds contribute equally to the momentum
injection rate ($\eta_{p}=2$), neglecting supernovae. $\tau$ is the
optical depth of IR radiation, weighted by the momentum of each
photon, which we conservatively take to be unity. $\delta v$ is the
width of the gas velocity distribution. Adopting the core CO line
widths of 100~km/s, we find $\dot{E}(SF) = 1.3 \times
10^{7}$~\Lsun. Thus, given uncertainties in $\tau$ and $\eta_{p}$, a
SF of 2.1\Msun/yr could sustain the energetics of the derived shocks
indefinitely.

The mechanical energy injected by jets from the central radio source
has also been estimated assuming the central source is an AGN (Nyland
et al., in prep). The implied jet power is $2.5 - 14 \times
10^{8}$~\Lsun, more than sufficient to power the shocks. Thus we find
SF, an AGN, or the combination of the two, can drive the observed
shocks.

\subsection{Dual PDRs}
A dual PDR is not able to simultaneously match the \twc SLED and
TIR. Relaxing the constraint of $\beta_{CO}$, the CO SLED can also be
fit with two PDRs, representing the high density core and envelope at
100~pc, with a reduced $\chi^2 = 3.3$. This highlights that the CO
SLED of NGC~1266 alone cannot unambiguously distinguish PDRs from
shocks.

H$_2$ emission also fails to distinguish a dual PDR from a PDR+Shock,
with the total H$_2$ intensity relative to CO roughly twice that
observed in both models. Similarly, the individual H$_2$ line ratios
relative to total observed CO are similar and do not break the
degeneracy between linear combinations of PDRs or Shocks and PDRs.

\subsection{ULIRG Physical Conditions}
The high IR-surface-brightness and gas densities found within the core
of NGC~1266 are closely analogous to the physical conditions of many
ULIRGs. 

Like many ULIRGs, NGC~1266 possesses an outflow in ionized and
molecular gas. In ULIRGs with outflows \water absorption lines are
blue-shifted by 100s km/s relative to those in emission
(i.e. Mrk231,\citealt{vanderWerf2010}). Our data show broad
[\ion{C}{2}] emission ($\sigma_{v} = 264$~km/s), but \water and CO
line centers are within 40~km/s of each other, implying a stronger
connection between \water and \twc in NGC~1266 than in ULIRGs.

The lower than average \thc/\twc in NGC~1266 is also a characteristic
of LIRGs and ULIRGs \citep{Aalto1995}, and indicates a low CO
optical depth generally attributed to a highly turbulent medium. We
speculate that in NGC~1266, a similar molecular gas velocity structure
is the result of shocks. Unlike some ULIRGs, this shocked material is
unlikely the result of a merger, as no extended \ion{H}{1} emission
has been detected \citep{Alatalo2011}. A high density is corroborated
by molecular line ratios, such as HCN/\thc which is an order of
magnitude larger than those observed in other early-type galaxies
\citep{Crocker2012}, and also similar to ULIRGs.

The infrared energy distribution observed with Spitzer and Herschel
indicates the dust in NGC~1266 is heated by a radiation field with
$G_0 = 20$. Assuming a spherical distribution of gas and dust with
$N{\rm (H)}/A_{\rm V} = 5.3 \times 10^{-22}~{\rm mag~cm^{2}}$, a
radius of 60~pc, and $\tau({\rm Si})/A_{\rm V} = 20$ (Chiar \&
Tielens, 2006), we find $A_{\rm V} \sim 10^3 {\rm mag}$, $A_{\rm 8
  \micron}$=38, and $\tau({\rm Si}) \approx 50$ to the center of
NGC~1266. This is much larger than observed, and whichever physical
process energizes the IR continuum, the observed (${\rm \approx
10s~\micron}$) continuum emission cannot originate at the center of
NGC~1266 behind the highest possible column.

\section{Conclusions}
The dominant energetic source of gas excitation within NGC~1266 is
challenging to determine. The extreme column densities of dust are
capable of hiding otherwise unambiguous diagnostics of AGN or star
formation. SPIRE-FTS observations of mid-J CO and \water emission now
provide the strongest evidence to date that the molecular gas of
NGC~1266 is energized by a $\approx$\vShock\ C-shock with a pre-shock
density of $\approx$\nShock. A PDR produces the low-J CO emission,
amounting to \fLowJ of the observed CO luminosity. This may originate
from the diffuse envelope surrounding the 60~pc core.


The discovery of shock-excited \water sets NGC~1266 and NGC~6240 apart
from other galaxies studied in detail. Such strong emission has only
been detected in extreme environments, i.e. Arp220
\citep{Rangwala2011}, Mrk231 \citep{vanderWerf2010}. Unlike these
other galaxies, NGC~1266 and 6240 lack emission from line transitions
above 300~K (Figure \ref{fig:COSLED} top-right), a signature of IR
pumping, leaving shocks as the dominant mechanism. \water emission has
proved to be a useful diagnostic to separate PDR and shocks, and of
the galaxies compared here, NGC~1266 has the highest observed
L(\water)/L(TIR) ratio.

The CO line-to-continuum ratio \citep{Meijerink2013} is a robust
diagnostic of shock excitation. We propose that [CI], which scales
linearly with L(CO) over $\approx$2~dex, is also a powerful shock
diagnostic. Another potential diagnostic is CO J=(7-6); like [CI], its
ratio to total CO is nearly constant for all the galaxies in
Figure~\ref{fig:COSLED}. The utility of these lines will be explored
further in our analysis of the entire BTP sample.

We conclude that the energetics of star-formation and AGN can both drive
the observed shocks. A star formation dominated energy source would
explain the lack of high ionization potential IR emission lines
usually detected in AGN, but not seen in NGC~1266 (i.e. [Ne~V],[O~IV]
\citealt{Smith2007}).

Despite having an infrared luminosity more than 30 times less than
that of a ULIRG, the compact molecular gas core in NGC~1266 has
strikingly similar high gas surface density, including
high-excitation, shocked gas and strong outflows.  Thus it is possible
for the bulk of a galaxy's gas reservoir to obtain the same extreme
physical characteristics of a ULIRG, even within an otherwise
quiescent stellar-dominated system.

{\acknowledgments We thank the referee for their valuable contributions
to the work presented. Research supported by NASA/JPL RSA
1427378(E.W.P., J.D.S. and A.F.C.), NASA JPL/Caltech 1426973(M.G.W.),
DAGAL network from the People Programme (Marie Curie Actions) of the
EU Seventh Framework Programme FP7/2007-2013/ under REA grant
agreement-number PITN-GA-2011-289313 (J.H.K).}

\end{document}